\documentclass[9pt,conference]{IEEEtran}
\usepackage{cite}
\usepackage{amsmath,amssymb,amsfonts}
\usepackage{algorithmic}
\usepackage{graphicx}
\usepackage{textcomp}
\usepackage{xcolor}
\usepackage[hyphens]{url}
\usepackage{fancyhdr}
\usepackage{hyperref}

\usepackage{comment}
\usepackage{color}
\usepackage{paralist} %
\usepackage{booktabs}
\usepackage{multirow}
\usepackage{soul}
\usepackage{listings} %
\usepackage{bbding} %
\usepackage[linesnumbered,ruled,lined,noend]{algorithm2e}

\usepackage{tikz}
\usepackage{todonotes} %
\usepackage{diagbox}

\newcommand{\hide}[1]{}

\newcommand{\bit}{\begin{compactitem}}
\newcommand{\eit}{\end{compactitem}}
\newcommand{\ben}{\begin{compactenum}}
\newcommand{\een}{\end{compactenum}}

\newif\ifsubmit
\submittrue
\ifsubmit
\newcommand{\rv}[1]{{#1}}
\else
\newcommand{\rv}[1]{{\color{blue}#1}}
\fi

\newif\ifarxiv
\arxivfalse
\ifarxiv
\newcommand{\oursystem}[0]{{TBA}}
\else
\newcommand{\oursystem}[0]{{SSDTrain}}
\fi

\title{\oursystem{}: An Activation Offloading Framework to SSDs for Faster Large Language Model Training}
\author{
    \IEEEauthorblockN{Kun Wu\IEEEauthorrefmark{1}\IEEEauthorrefmark{2}\IEEEauthorrefmark{6}, Jeongmin Brian Park\IEEEauthorrefmark{1}\IEEEauthorrefmark{2}\IEEEauthorrefmark{6}, Xiaofan Zhang\IEEEauthorrefmark{1}\IEEEauthorrefmark{3}, Mert Hidayeto\u{g}lu\IEEEauthorrefmark{4}, Vikram Sharma Mailthody\IEEEauthorrefmark{2},
    \\Sitao Huang\IEEEauthorrefmark{5}, Steven Sam Lumetta\IEEEauthorrefmark{6}, Wen-mei Hwu\IEEEauthorrefmark{2}\IEEEauthorrefmark{6}}
    \IEEEauthorblockA{\IEEEauthorrefmark{2}NVIDIA \IEEEauthorrefmark{3}Google \IEEEauthorrefmark{4}Snowflake \IEEEauthorrefmark{5}University of California, Irvine \IEEEauthorrefmark{6}University of Illinois Urbana-Champaign
    \\\IEEEauthorrefmark{1}The three authors made equal contribution. Corresponding author is Kun Wu, kunw@nvidia.com.}
}

\begin{document}
\maketitle

\begin{abstract}
The growth rate of the GPU memory capacity has not been able to keep up with that of the size of large language models~(LLMs), hindering the model training process. In particular, activations---the intermediate tensors produced during forward propagation and reused in backward propagation---dominate the GPU memory use. This leads to high training overhead such as high weight update cost due to the small micro-batch size. To address this challenge, we propose \oursystem{}, an adaptive activation offloading framework to high-capacity NVMe SSDs. \oursystem{} reduces GPU memory usage without impacting performance by fully overlapping data transfers with computation. \oursystem{} is compatible with popular deep learning frameworks like PyTorch, Megatron, and DeepSpeed, and it employs techniques such as tensor deduplication and forwarding to further enhance efficiency. We extensively experimented with popular LLMs like GPT, BERT, and T5. Results demonstrate that \oursystem{} reduces 47\% of the activation peak memory usage. \rv{Meanwhile}, \oursystem{} perfectly overlaps the I/O with the computation and incurs negligible overhead. Compared with keeping activations in GPU memory and layerwise full recomputation, \oursystem{} achieves the best memory savings with negligible throughput loss. We further analyze how the reduced activation memory use may be leveraged to increase throughput by increasing micro-batch size and reducing pipeline parallelism bubbles.
\end{abstract}

\section{Introduction}
\label{sec:intro}

LLMs now drive a wide range of applications, including chatbots~\cite{openaiChatGPT2022}, search~\cite{BingChatMicrosoft2023}, content generation~\cite{midjourneyMidjourney2022}, reasoning~\cite{langchainLangChain2022}, etc. These models, when sufficiently large in size, demonstrate emergent abilities~\cite{weiEmergentAbilitiesLarge2022} and thus the capability of handling complicated tasks. Such a phenomenon drives model designers to continue to scale up the size of LLMs, carrying more parameters. The already formidably high training costs continue to grow: training \mbox{GPT-4}, for example, cost US\$100 million, a 21$\times$ increase over training~\mbox{GPT-3}~\cite{davidmeyerCostTrainingAI2024}.

\begin{figure}[!t]
\centering{\includegraphics[width=\linewidth]{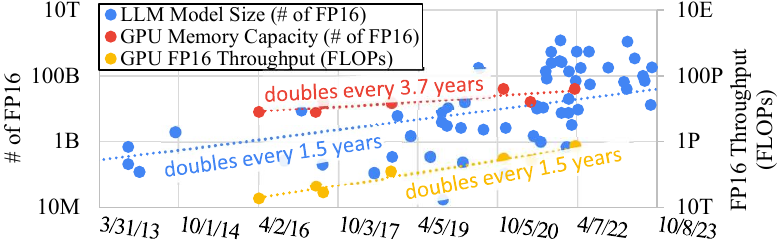}\vspace{-5pt}}
\caption{\label{fig:trend_scale} The growth of FP16 throughput of GPUs for deep learning training is aligned with the model size of LLMs, but GPU memory capacity falls behind~\cite{theepochaiAnnouncingEpochAI2023}. Horizontal axis shows release date.  Points represent both Nvidia 100-level GPUs since K100 and Google TPUs.}
\end{figure}

GPU memory capacity has become a bottleneck for the continued growth of LLMs.
As Fig.~\ref{fig:trend_scale} shows, the increase of GPU memory capacity is around 60\% slower than the LLM size scaling speed and the GPU FP16 throughput improvement. About 80\% of the GPU memory used to train recent LLMs consists of activations~\cite{liuWinnerTakeAllColumnRow2023,korthikantiReducingActivationRecomputation2022}, the intermediate tensors produced by forward propagation and reused in backward propagation.
Furthermore, the memory needed for activations is growing more rapidly than any other memory use, making GPU memory a more serious constraint for future LLM training~(see Sec.~\ref{sec:llm_scaling} for details).

Common mitigations are to reduce batch size or through gradient accumulation.  With gradient accumulation, a batch is divided into micro-batches that are %
processed separately between gradient updates.  Although gradient accumulation has been adopted by many LLMs~\cite{jiangMegaScaleScalingLarge2024,shoeybiMegatronLMTrainingMultiBillion2020a,workshopBLOOM176BParameterOpenAccess2023}, the GPU computation stack is not designed for small inputs, and %
both mitigations lead to device under-utilization~\cite{DissectingBatchingEffects,anthonyCaseCoDesigningModel2024} and suboptimal math library performance~\cite{aminabadiDeepSpeedInferenceEnabling2022}. 
Intuitively, %
a smaller batch size might reduce total training computation through faster convergence. However, LLM trainers have identified a %
critical batch size to each model, below which convergence speed increases negligibly or even decreases~\cite{kaplanScalingLawsNeural2020,mccandlishEmpiricalModelLargeBatch2018}.  Notably, critical batch size grows %
during training, as training loss is reduced.
Another common approach to reducing GPU memory use is activation checkpointing.  With this, only some activations are kept in GPU memory, while others are flushed and then recomputed during backward propagation.  For an $L$-layer model, activation checkpointing reduces memory requirements from $O(L)$ to $O(\sqrt{L})$~\cite{chenTrainingDeepNets2016}.  However, as Sec.~\ref{sec:llm_scaling} shows, even this alone is insufficient to eliminate the bottleneck posed by GPU memory limits for future LLMs.

This work proposes \oursystem{}, a software framework that offloads activations to NVMe SSDs and reloads activations just before they are needed in backward propagation.  \oursystem{} is able to overlap activation transfers fully with computation, thereby reducing activation memory usage without incurring significant performance overhead. SSDs are a more attractive target than main~(CPU) memory for several reasons. First, clusters and cloud instances~\cite{microsoftNDA100V4series2024,googleGPUMachineTypes,ncsaDeltaProjectProfile} are typically limited in host memory capacity (100--250 GB/GPU), while SSDs offer much higher capacity. The limited host memory is also consumed by input, metadata, etc., which further reduces the amount of memory available for activation offloading. Second, host memory bandwidth is shared across training management tasks and offloaded computation~\cite{kamahoriFiddlerCPUGPUOrchestration2024,renZeROOffloadDemocratizingBillionScale2021,songPowerInferFastLarge2023} running on the host CPU and can be quite limited and even unpredictable ~\cite{baeFlashNeuronSSDEnabledLargeBatch2021} for saving and restoring activations. In contrast, the SSD bandwidth can be dedicated to the activation offloading during training. %
Third, SSDs are more elastic, both %
by adding more SSDs and even PCIe switches if necessary---as well as through the use of optional remote high-throughput storage~\cite{googleGoogleCloudHyperdisk,lockwoodArchitecturePerformancePerlmutter2024}. \rv{This} allows data centers to keep up with the fast-growing size of activations. In contrast, the memory capacity of GPU cloud instances and cluster nodes is much more difficult to extend. %

This work makes the following main contributions. 
\begin{enumerate}[1.]
\item We design and implement the \oursystem{} framework to offload LLM activations to NVMe SSDs. We demonstrate the viability of \oursystem{} on large-scale systems by modeling the performance, SSD lifespan and the required per-GPU PCIe bandwidth. 
\item With all code in Python except for a tiny CUDA API hooking library, \oursystem{} works with the latest PyTorch, and distributed frameworks. We developed and tested \oursystem{} with Megatron-DeepSpeed~\cite{microsoftMicrosoftMegatronDeepSpeedOngoing2019} on a 2-GPU node with 7$\times$ Intel Optane SSDs. 
\item Evaluation shows \oursystem{} \rv{matches the original system's} training time while reducing the activations peak memory use by up to 47\%. This \rv{proves} that \oursystem{} overlaps the data transfer fully with computation. Compared with keeping activations and layerwise full recomputation, \oursystem{} \rv{obtains} the best performance and the least memory peak.  We further analyze how the reduced activation memory use may increase throughput by increasing micro-batch size and reducing pipeline parallelism bubbles.

\end{enumerate}

\noindent
\rv{The code repository is public at \url{https://github.com/K-Wu/FlashTrain}.}

\section{Background}

\subsection{Transformer-Based LLM} %
Most LLM architectures, including GPT~\cite{radfordLanguageModelsAre2019},  are transformer-based~\cite{vaswaniAttentionAllYou2017}. These models consist mainly of transformer layers. 
Each transformer layer is primarily made up of an attention block and a multi-layer perception (MLP) block. 
GPT is a decoder-only model because it only involves transformer decoder layers. A\rv{n} encoder layer has the same structure as the decoder layer except that the latter imposes causality on the attention mask.
\rv{T}ransformer models are classified as (1) encoder-only, e.g., BERT~\cite{devlinBERTPretrainingDeep2019}, (2) decoder-only, and (3) encoder-decoder, e.g., T5~\cite{raffelExploringLimitsTransfer2023}. In encoder-decoder models, decoder layers take in both outputs from the encoders and another text and apply two attention blocks---the self-attention block is applied to the new text, and the cross-attention block is applied among the tokens in the sequence from the encoder and tokens in the new text.

Parallelizing LLM training involves partitioning and/or replicating the model and the data into different GPUs~\cite{xuGSPMDGeneralScalable2021}. Pipeline parallelism (PP), data parallelism (DP), and \rv{tensor} parallelism (\rv{T}P) are the three \rv{widely-adopted} levels of parallelism available to all LLM models.
PP \rv{divides} the model \rv{ and places} chunks of layers on different GPUs.  In a step, when the GPUs finish their layers, the output is passed to the GPUs owning next layers.
DP replicates the models in different groups of GPUs and assigns separate micro-batches to each group.
\rv{T}P shards a weight tensor and puts shards onto different GPUs. Each GPU performs a portion of the computation using its shard for the corresponding operator.
Zero Redundancy Optimizer (ZeRO)~\cite{rajbhandariZeROMemoryOptimizations2020a} further reduce memory use with DP by sharding the optimizer states, and/or optionally the gradients and parameters across these GPUs.

\subsection{GPU Memory Capacity and Model Throughput}
\label{sec:llm_scaling}
As Fig.~\ref{fig:eval_dse} of Sec.~\ref{sec:evaluation} will show, the GPU memory capacity is limiting the model throughput. By offloading the activations to SSDs, \oursystem{} can alleviate this limitation and improve the per-GPU model throughput. An important question is if the GPU memory capacity will continue to be the limiting factor of per-GPU model throughput according to the trend of LLM scaling. This section shows that the historical trend will make GPU memory capacity an even more important limiting factor of the per-GPU model throughput.

Neural scaling laws~\cite{jordanhoffmannTrainingComputeOptimalLarge2022,kaplanScalingLawsNeural2020,mccandlishEmpiricalModelLargeBatch2018} guide LLM scaling as computing power increases. We follow these laws in our reasoning.
The whole-system GPU compute throughput $C\propto ND_{batch}$, where $N$ is the number of parameters and $D_{batch}$ is the number of tokens in a batch~\cite{brown2020languagemodelsfewshotlearners}. Chinchilla Scaling ~\cite{jordanhoffmannTrainingComputeOptimalLarge2022} concludes that the optimal model design follows $N\propto C^{0.5}$, which implies $D_{batch}\propto C^{0.5}$ to saturate the GPU throughput.  Whole-system GPU memory use consists of two parts: activations, which require $S_{activations}\propto \frac{N}{h}D_{batch}$, where $h$ is the hidden dimension in the layers and is a slow growing function of $N$, e.g., $h\propto N^{1/3}$, and all other memory use, $S_{others}\propto N$, including parameters, gradients, and optimizer states. Comparing the factors, we can deduce that (1) $S_{activations}$ grows faster than $S_{others}$, and (2) whole-system memory use, which is dominated by the activations, grows %
slightly slower than the compute throughput $C$~(approximated $C^{5/6}$).
However, Fig.~\ref{fig:trend_scale} shows that the GPU memory capacity historically grows~(red dotted line) at 41\% the growth rate of the compute throughput~(yellow dotted line). Therefore, \textbf{GPU memory capacity will become increasingly inadequate for saturating the compute throughput, and memory for activations will continue to dominate the GPU memory usage.}

What about activation checkpointing? Revisiting the prior equation, $S_{activations}\propto \frac{N}{h}D_{batch}\propto LhD_{batch}$ where $L$ is the number of layers. Checkpointing makes the new activations memory use $S_{activations}^\prime \propto \sqrt{L}hD_{batch}$. Since $L$ and $h$ grow when $N$ increases and $D_{batch}\propto C^{0.5}$, $S_{activations}^\prime$ still grows faster than $S_{others}$.

\subsection{SSD Endurance}

Trends in price, latency, and bandwidth have led to the widespread adoption and integration of SSDs into cloud instances and clusters~\cite{microsoftNDA100V4series2024,googleGPUMachineTypes,ncsaDeltaProjectProfile}.
Flash's random write latency is reduced to tens of microseconds~\cite{samsungUltraLowLatencySamsung2017}, and NVMe SSD data rates are now a few GB/s.

SSD endurance remains a concern: how long will SSDs last in the write-intensive activation offloading?
SSD endurance is determined by the type and numbers of cells, write amplification factor~(WAF), and over-provisioning.
SSD cells can be purposed to store one bit or multiple levels.
Generally, the more bits a cell stores, the shorter its lifetime in program-erase~(PE) cycles. WAF is the ratio of media write amount to host write amount---SSD writes pages at a time but erases blocks of pages, a coarser granularity. Erasing a partially empty block requires that remaining valid pages be relocated, causing write amplification. 
In turn, vendors adopt over-provisioning to reserve some blocks for wear leveling, evening out the writes across blocks.

Notably, SSD endurance rating uses the JESD testing method~\cite{jedecsolidstatetechnologyassociationJESD218BSolidStateDrive2016} which performs random writes after tough preconditioning. In our scenario, the writes are large and sequential as each tensor being offloaded is easily hundreds of MBs in size. Such writes are more endurance-friendly compared with the writes used to determine the JESD rating. 
For example, \mbox{3-DWPD} SSDs generally allow about 2.5$\times$ as many sequential writes than expected from the JESD rating~\cite{lenovoWhatNeedKnow2023,qnapsystemsinc.QNAPNASSolution2108, smartmodulartechnologiesinc.WhySMARTOverProvisioning2024}. Vendor guidelines~\cite{solidigmSolidigmSSDEndurance,intelOverProvisioningNANDBasedIntel2018,samsungOverProvisioningBenefitsSamsung2019} and empirical data~\cite{maneasOperationalCharacteristicsSSDs2022} corroborate this difference.
Sec.~\ref{sec:projected_life} uses modeling to demonstrate why mainstream data center SSDs are viable options to support \oursystem{} deployment in a large-scale LLM training system.

\begin{figure}[!t]
\centering{\includegraphics[width=0.95\linewidth]{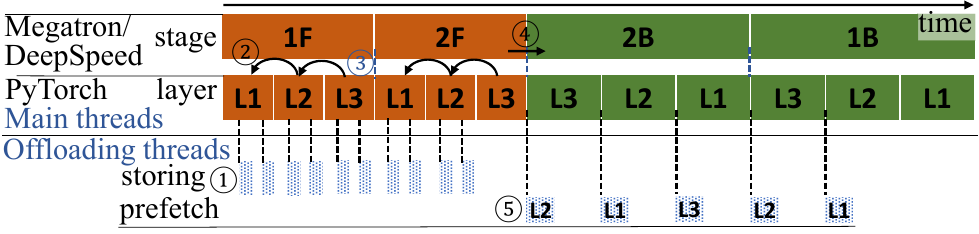}\vspace{-5pt}}
\caption{\label{fig:pipe}\oursystem{} timeline of a step of a 2-microbatch 3-layer~(L) model. }
\end{figure}

\begin{table}[!t]
\centering
{\footnotesize
\caption{Comparing \oursystem{} to other LLM systems with activation offloading features~\cite{shengFlexGenHighThroughputGenerative2023,alizadehLLMFlashEfficient2024,rajbhandariZeROinfinityBreakingGPU2021}. Without backward propagation, inference systems may discard intermediate tensors once a layer is done. We generalize ``\textbf{Activation}'' to refer to key-value~(KV) cache as well because it is reused across steps. \label{tab:salesman}\vspace{-5pt}}
\begin{tabular}{ @{}l@{\hspace{\tabcolsep}}l@{\hspace{0.5\tabcolsep}}c@{\hspace{0.5\tabcolsep}}c@{\hspace{0.5\tabcolsep}}c@{\hspace{0.5\tabcolsep}}||c@{}}
\toprule
    &       & \rotatebox{60}{Flexgen} & \rotatebox{60}{\parbox{1cm}{LLM in a Flash}} &\rotatebox{60}{\parbox{1cm}{ZeRO-Infinity}} & \rotatebox{60}{\textbf{\oursystem{}}} \\ 
\midrule  
\multicolumn{2}{@{}l}{\textbf{Training}}                           &           &  & \Checkmark & \Checkmark \\\cline{1-2}
\multirow{2}{*}{\textbf{\begin{tabular}[c]{@{}l@{}}Activation\\ offloading\end{tabular}}}        & \textbf{to main memory}        &    \Checkmark       &  \Checkmark &   Checkpoints only & \Checkmark \\\cline{2-2}
                                   & \textbf{to SSD}          &  \Checkmark         &  &   & \Checkmark \\\cline{1-2}
\multicolumn{2}{@{}l}{\textbf{Direct GPU--SSD data path}}                &           &  &  & \Checkmark\\\cline{1-2}
\multicolumn{2}{@{}l}{\textbf{Async data transfer}}                &           &  &  & \Checkmark\\\cline{1-2}
\multicolumn{2}{@{}l}{\textbf{Interoperability}} & & & & \Checkmark\\
\bottomrule
\end{tabular} 
}
\end{table}

\subsection{SSD Offloading Systems for LLM}
GPUDirect Storage~(GDS) enables a direct data path between GPU and NVMe SSDs~\cite{inupakutikaQuantifyingPerformanceGains2022,rajbhandariZeROinfinityBreakingGPU2021,yangProTrainEfficientLLM2024}, removing the need for a CPU bounce buffer to enhance bandwidth and reduce both latency and CPU load. 

To mitigate the training overhead caused by the GPU memory capacity limit, SSDTrain has three key differences with related work~\cite{sunSTRONGHOLDFastAffordable2022,yangProTrainEfficientLLM2024}: SSDTrain offloads (a) activations to (b) the SSDs (c) with negligible performance overhead. To the best of our knowledge, SSDTrain is the first work that leverages SSD to offload activations for LLM training.
Table~\ref{tab:salesman} illustrates some other \oursystem{}'s features:

\noindent
\textbf{Direct \rv{and async} GPU--SSD data \rv{transfer}.} As Sec.~\ref{sec:intro} \rv{shows}, transfer via CPU \rv{impacts} efficiency. 
\rv{Besides, existing} systems either block the training when loading offloaded data, or synchronize at each layer. \rv{Thus}, the I/O latency is in the critical path. \oursystem{} hides the I/O latency by overlapping I/O with computation. 

\noindent
\textbf{Interoperability.} Since LLM training requires a synergy of Python packages and the ecosystem is rapidly evolving, it is vital for offloading to have good interoperability with other components. \oursystem{} logic is local to processes and can work with distributed frameworks, e.g., Megatron. In contrast, DeepSpeed's offloading features, e.g., ZeRO-Infinity, are available only in certain ZeRO stages. ZeRO stage determines what is sharded. For example, stage-3 ZeRO in Fig.~\ref{fig:projected_perf_model} sharded optimizer states, gradients, and weights across the GPUs.

\section{Design and Implementation}

\subsection{Overview of the \oursystem{} Framework}
\oursystem{} implements a {\it tensor cache} to manage the offloading and reloading of tensors, facilitating the release of memory as well as the prefetch of tensors back to memory before they are needed for backward propagation.
Fig.~\ref{fig:pipe} exemplifies how \oursystem{} works.
\oursystem{} launches its own threads to store tensors (\textcircled{1}) and to reload them~(\textcircled{5}). In forward propagation (F), offloading of an activation starts once the operator producing it finishes~(\textcircled{1}). When activations are reused in backward propagation (B), prefetching~(\textcircled{5}) occurs in the reverse order of layers as recorded during forward propagation~(\textcircled{2}). If the last layer begins backward propagation immediately after its forward propagation (L3 in micro-batch 2 in the example), its activations are kept~(\textcircled{4}). \oursystem{} keeps individual records for each micro-batch. Upon micro-batch changes~(\textcircled{2}), \oursystem{} switches its own record to the one corresponding to the new micro-batch. %

Fig.~\ref{fig:software_arch} shows the \oursystem{} workflow. SSDTrain retrieves the amount of computation and activation size of the model from the model instance, GPU throughput, and SSD bandwidth. Then, SSDTrain sets the activation offload amount accordingly.
The tensor cache manages the activations and performs tensor offloading and loading. To achieve this, it uses PyTorch hooks to alter PyTorch execution. Sec.~\ref{sec:tensor_cache} details the design and implementation of the tensor cache. \oursystem{} has the SSD offloader that targets NVMe SSDs within the same node and the CPU offloader that targets host memory. Each offloader encapsulates the logic to transfer CUDA tensors to and from a target. The SSD offloader leverages the GDS python binding, kvikio~\cite{nvidiaRapidsaiKvikioKvikIO2022}. Using the \texttt{LD\_PRELOAD} interposition mechanism, the CUDA malloc hook library alters CUDA memory allocation and free API calls so that the memory is properly registered and deregistered for best GDS performance. This allows us to keep the PyTorch memory allocator for easy comparison with the baseline, without replicating its implementation in a PyTorch pluggable memory allocator or modifying the PyTorch C++ code. The CPU offloader is for future work on clusters with massive remote SSD storage. It is backed by an allocator with pre-allocated host-pinned memory. The pool size is determined by profiling the first training step. Hints are added to Megatron's and DeepSpeed's schedulers, e.g., \textcircled{3} and \textcircled{4} in Fig.~\ref{fig:pipe}. E.g., for DeepSpeed's scheduler, hints are added before and after the execution of each command, e.g., computing the micro-batch \texttt{i}, communication so that the tensor cache gets notified about the upcoming stage and the completion of an action. Accordingly, the tensor cache can prefetch data, or wait for I/O to complete.

To use \oursystem{}, a few lines are to be added in the existing script: They register the PyTorch hooks, bookkeep the weights to not offload them, and monkey-patch~\cite{wikipediaMonkeyPatch2024} the schedulers. %
 
\oursystem{} extends naturally to distributed settings such as use with ZeRO, because frameworks such as DeepSpeed and Megatron divide the workload into processes built on top of PyTorch's built-in tensor functionality.
By working below PyTorch and keeping each process' activities local, \oursystem{} applies directly to distributed launches.

\subsection{Hook-Based Implementation of Tensor Cache}
\label{sec:tensor_cache}

To benefit from tensor offloading, the GPU memory that the offloaded tensors own must be released when the tensors are not in use. However, PyTorch by default stores a reference to all the activations on the computational graph, disallowing the GPU memory to be reclaimed. The tensor cache alters the PyTorch execution so that the identifiers of the activations are registered on the computation graph; upon PyTorch's reusing the activation tensor, the tensor cache uses the identifier from the computational graph as the key to return the requested tensor. In the forward propagation, when the tensor finishes offloading, the tensor cache no longer holds a reference to it, allowing its memory to be reclaimed by Python garbage collection once the control flow gets out of the function scope where the tensor object is used. In the backward propagation, the tensor cache holds a reference to the tensor by loading it from the SSD before its use; when all the module scopes the tensor is referred to have been finished, the reference is no longer held, allowing its memory to be reclaimed.
In short, the tensor cache is the in-memory structure that manages the references to all activations and tracks activations' states, including if they are being offloaded, the path in the file system, etc.

Tensor cache uses PyTorch hooks to alter its execution behavior.
The forward hook pair works in the forward propagation: The start of a module triggers the forward pre hook, and the finish of a module triggers the forward hook. 
Tensor cache maintains the current scope stack using the forward hook pair: Upon entrance to a module, the module is pushed to the stack;
Upon module exit, it is popped out.

Backward hook pair is similar. 
When entering a module, the tensor cache prefetches activations in upcoming modules. Sec.~\ref{sec:prefetch} details prefetching.
When exiting a module, the tensor cache removes it from the scope lists of all activations. Activations no longer in use are removed, whose memory will be released by garbage collection.

\begin{figure}[!t]
\centering{\includegraphics[width=0.7\linewidth]{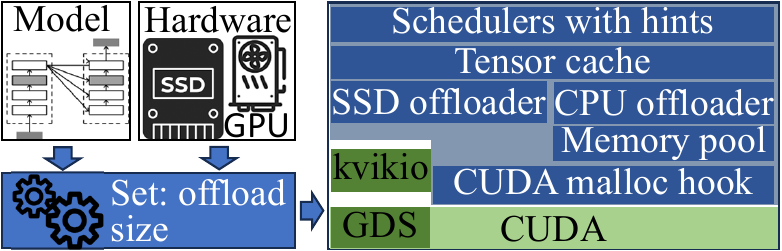}\vspace{-5pt}}
\caption{\label{fig:software_arch} \oursystem{} workflow. \oursystem{} components are shown as blue blocks.}
\end{figure}

When a tensor is to be registered onto the computation graph, the pack hook is called to produce a value to be registered instead.
When the tensor is reused, the unpack hooks is called to take in the object on the computation graph and return the original tensor.
Fig.~\ref{fig:hooks} illustrates tensor cache's activity when pack or unpack hook is triggered.
When the multiply operator $\texttt{x}\cdot\texttt{w}$ finishes (\textcircled{1}), the pack hook is called (\textcircled{2}) on the input \texttt{x} and weights \texttt{w}.
Tensor cache has a record of weights, and accordingly returns \texttt{w} to let it be registered on the graph as is.
The tensor will also be returned as is if the tensor is on CPU or it is too small (Line 2 in Alg.~\ref{algo:tensor_cache_hooks}).
As line 6 in Alg.~\ref{algo:tensor_cache_hooks} shows, the tensor cache does not offload tensors but only keeps a record when the module is to be kept in the memory or in backward propagation.
The first condition holds true when the amount of activation in this step before the current tensor reaches the size set in Fig.~\ref{fig:software_arch}.
The second condition is true when an activation-checkpointing-enabled function does recomputation in the backward propagation to reproduce the activations.
For tensor \texttt{x} in Fig.~\ref{fig:hooks}, the tensor cache stores it to the SSDs (\textcircled{3}), updates the amount of activations offloaded in this step, and returns a tensor identifier.
When the unpack hook is triggered (\textcircled{B}), in the backward propagation (\textcircled{A}), the tensor cache either waits until the prefetch finishes(\textcircled{C}), and eventually returns the tensor. 

\subsection{Tensor Cache Mechanisms and Optimization}
\subsubsection{Deduplicating Tensors and Excluding Weights}

Tensor cache has a \texttt{get\_id()} function to assign a unique identifier to each tensor.
The shortcoming of PyTorch native \texttt{id()} is that its returned value is related to the GPU memory address. As \oursystem{} offloads activations, the latter will be cleared by garbage collection once the control flow goes out of its use scope.
The GPU memory address may be reused, causing identifier collision. 
To solve this, \texttt{get\_id()} combines the timestamp when it first processes the tensor with the tensor shape as the unique identifier: When \texttt{get\_id()} processes a tensor \texttt{t} for the first time, \texttt{get\_id()} adds the current timestamp as an additional attribute to the tensor's underlying storage \texttt{t.untyped\_storage()} instead of \texttt{t}.
This is because sometimes PyTorch creates new \texttt{torch.Tensor} objects representing the identical tensor.
All future \texttt{get\_id()} calls get the attribute value.
This deduplicating scheme helps prevent redundant I/Os.

PyTorch registers all needed tensors in backward propagation into the computational graph, involving activations and weights. As this work focuses on activations, the tensor cache excludes the weights. To this end, before training, the tensor cache records the identifiers of all weights. As linear layers store the transpose of weights for backward propagation, the unique identifiers of the transpose are recorded. One benefit of our \texttt{get\_id()} scheme is that the identifier for the transpose of the same weights tensor remains consistent across steps. This is because the transpose uses the original tensor’s underlying storage, which we already assigned a timestamp to before training.

\subsubsection{Offloading and Forwarding Tensors}
\label{sec:prefetch}

The tensor cache has two thread pools---one for storing tensors and the other for loading tensors. Submitted jobs are executed in first-in-first-out (FIFO) order.

To hide the I/O latency, the tensor cache starts prefetching each activation before the corresponding module's backward propagation. 
The activations in the last module is kept in GPU memory so they need not be prefetched. %
This simple scheme suffices because in PyTorch, CPU submits GPU kernel launches and memory operations ahead of GPU execution. Prefetching schemes are equivalent as long as there are always I/O tasks in GPU job queue to keep PCIe busy.

\begin{figure}[!t]
\centering{\includegraphics[width=0.85\linewidth]{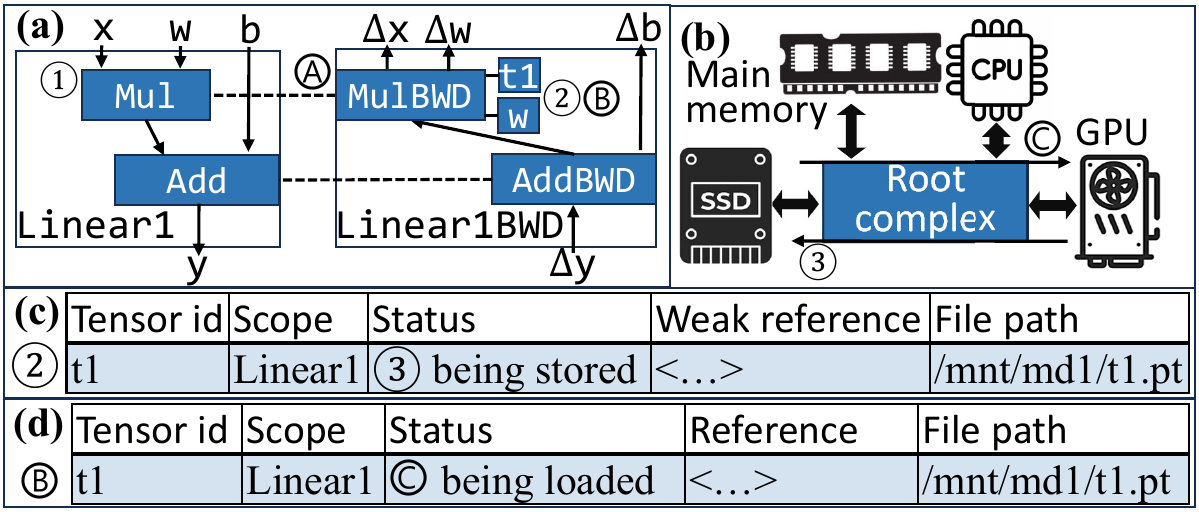}\vspace{-5pt}}
\caption{\label{fig:hooks} Tensor cache registers hooks to offload tensors and reload tensors. \textbf{(a)} shows the computational graph. \textbf{(b)} shows the hardware data path. \textbf{(c)} and \textbf{(d)} show the tensor cache state when the pack or unpack hook is triggered.  }
\end{figure}

Upon loading a tensor, if it is still being stored, the tensor cache will return its in-memory reference to skip loading from SSD. We call this data forwarding. \rv{E.g.}, in Fig.~\ref{fig:hooks}, when PyTorch retrieves \texttt{x} from the \texttt{MulBWD} node, if \texttt{x} is still being stored, it is in memory. Instead of loading the tensor, the tensor cache returns \rv{\texttt{x}'}s reference convert\rv{ed from} the weak reference and store the obtained reference in the tensor cache for future if it is used in other scopes.

\begin{algorithm}[!t]
{\footnotesize
\DontPrintSemicolon
    \KwIn{\rv{T}ensor cache \texttt{tc}, tensor \texttt{t}, and/or object to unpack \texttt{obj}.}
    
    \SetKwProg{Fn}{Function}{:}{}
    \SetKwFunction{FPCKH}{\texttt{pack\_hook}}
    \Fn{\FPCKH{\texttt{t}}}{
     \lIf{\texttt{tc.is\_weights(t) or t.is\_cpu or math.prod(t.size())<2**20}}{
     \Return{\texttt{t}}
     }

    \texttt{tid = get\_id(t)}

    \texttt{tc.add\_to\_current\_scope(tid)}

    \If{\texttt{tc.is\_offload\_amount\_reached() or tc.is\_current\_in\_backward()}}{
     \texttt{tc.keep\_in\_gpu\_memory(tid,t)}
     }\lElse{
     \texttt{tc.offload(tid,t)}
     }

    \Return{\texttt{tid}}
    }

    \SetKwFunction{FUPCKH}{\texttt{unpack\_hook}}
    \Fn{\FUPCKH{\texttt{obj}}}{
        \lIf{\texttt{isinstance(obj,torch.Tensor)}}{
            \Return{\texttt{obj}}
        }
        \texttt{tc.load\_or\_wait\_load(obj)} 
        
        \Return{\texttt{tc.get\_loaded\_tensor(obj)}}
    }
}
\caption{\small Pack--unpack hook pair used by the tensor cache.}
\label{algo:tensor_cache_hooks}
\end{algorithm}

\subsection{SSD Write Amount, Bandwidth, and Lifespan}
\label{sec:projected_life}
To confirm if our design is viable in large-scale systems, particularly concerning SSD endurance and required bandwidth, we conduct performance modeling to obtain forward propagation time per training step and the size of activations produced in the process. 
We extend the performance model package \texttt{llm-analysis}~\cite{liLLMAnalysisLatencyMemory2023}, which models the forward propagation of each transformer layer as a simple pipeline: 
$t= \max\left(\sum_{l}\max\left(t_{l,compute}, t_{l,memory}\right), t_{ZeRO,communicate}\right)$,
where $l$ denotes any layers inside a transformer layer. When ZeRO is enabled, the ZeRO communication time is assumed to be perfectly pipelined with the non-ZeRO operations at the transformer layer level. 

 We model the required PCIe write bandwidth per GPU as the total amount of activations divided by half the training time. %
 The lifespan is then projected as $t_{life} = S_{endurance}\cdot t_{step}/S_{activations}$ where $S_{endurance}$ is the lifetime writes allowed by the SSD endurance rating, $S_{activations}$ is the amount of activations per training step, and $t_{step}$ is the step time. We validated the $S_{activations}$ formula with experiments in Sec.~\ref{sec:evaluation}. We assume four Samsung 980 PRO 1TB for each GPU, and assume the WAF is 2.5 in JESD rating and 1 in our scenario. We also relax the data retention period: NAND flash gets 86$\times$ PE cycles when the data retention period is relaxed from 3 years to 1 days~\cite{caiFlashCorrectandrefreshRetentionaware2012,yucaiErrorPatternsMLC2012,liuOptimizingNANDFlashBased2012,kimBehemothFlashcentricTraining2021}.
With these, we obtain Fig.~\ref{fig:projected_perf_model}. We use measured data from Megatron-LM~\cite{shoeybiMegatronLMTrainingMultiBillion2020a}. GPUs are A100 PCIe. Among all cases, the projected lifespan is more than 2 years, and the write bandwidth per GPU is no greater than 12.1GB/s.
Moreover, when the system size and/or the model size scales up, the required PCIe write bandwidth reduces, and the projected lifespan increases.  This is because larger systems imply increased communication overhead and reduced  computation efficiency, slowing down training on GPUs. 

We also estimate the maximal activations size each GPU produces per step by assuming only two layers in a row are in GPU memory at the same time while all other activations are offloaded. Then, the activation maximal micro-batches produce in a step are shown as diamond marks in Fig.~\ref{fig:projected_perf_model}. The maximal activations size per GPU ranges from 0.4 TB to 1.8 TB, while the micro-batch size ranges from 8 to 32. Activations so large can no longer be held by the main memory and therefore SSD is the only choice as offloading target.

\section{Evaluation}
\label{sec:evaluation}

\subsection{Experimental Setup}
\label{sec:eval_method}
We use a machine with 2$\times$ A100 PCIe GPUs and 7$\times$ Intel P5800X SSDs, as Table~\ref{tab:configurations} specifies. The SSDs are organized into two RAID0 arrays: one with 3 SSDs, and the other with 4 SSDs. Each array is the dedicated  target of one A100. We measured the memory use of the A100 with 4 SSDs during evaluation. For consistency, the GPUs are locked at base frequency. The latest Megatron-DeepSpeed~\cite{microsoftMicrosoftMegatronDeepSpeedOngoing2019} is used, which incorporates DeepSpeed techniques into Megatron.

We measure the pretraining performance on BERT~\cite{devlinBERTPretrainingDeep2019} as an encoder-only model, GPT~\cite{radfordLanguageModelsAre2019} as a decoder-only model, and T5~\cite{raffelExploringLimitsTransfer2023} as an encoder-decoder model. 
We use the OSCAR dataset~\cite{ortiz-suarez-etal-2020-monolingual,OrtizSuarezSagotRomary2019}.

\begin{figure}[!t]
\centering{\includegraphics[width=0.85\linewidth]{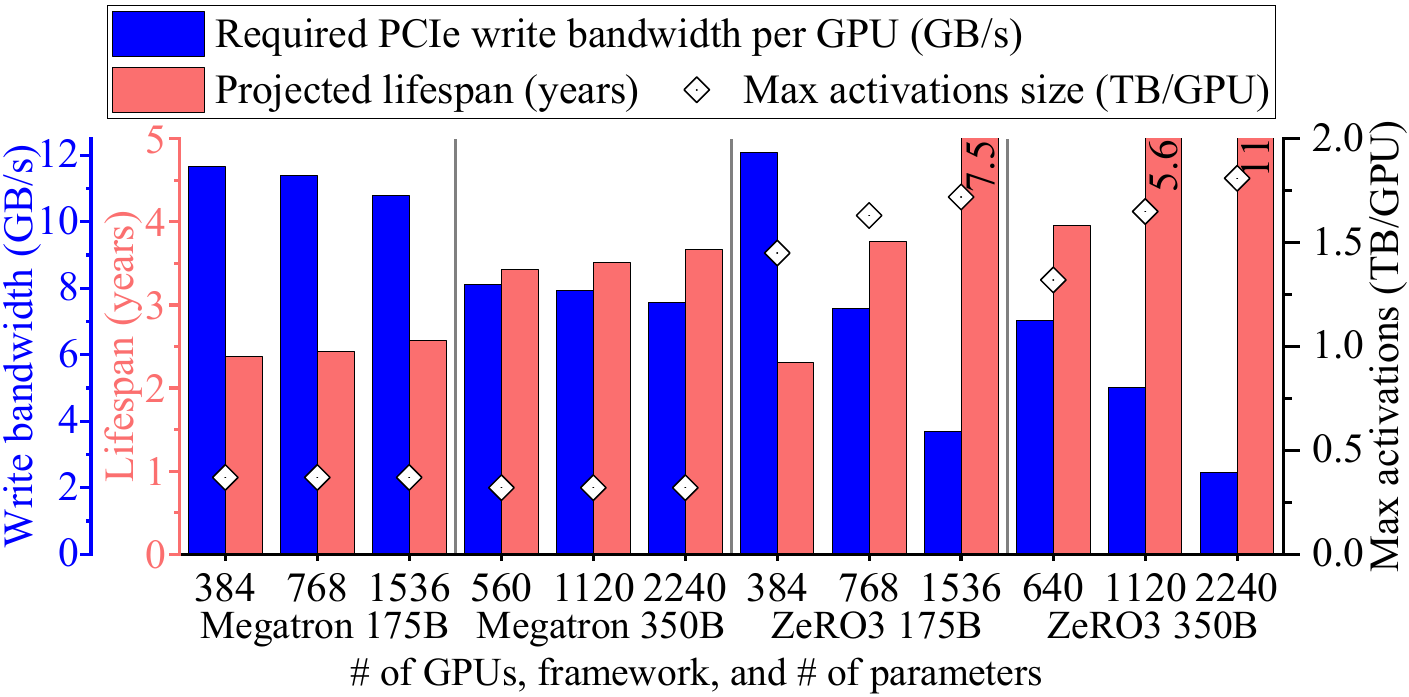}\vspace{-7pt}}
\caption{\label{fig:projected_perf_model} Estimate of SSD lifespan, PCIe write bandwidth and maximal activations size per GPU. Lifespans longer than 5 years are shown on top of the pink bars. ZeRO3 stands for DeepSpeed with stage-3 ZeRO.}
\end{figure}

\begin{table}[!t]
{\footnotesize
\caption{Evaluation system configuration.\vspace{-5pt}}\label{tab:configurations}
\begin{tabular}{rl}
    \toprule
    \textbf{CPU (Memory)} & 2$\times$ AMD EPYC 7702 64-core (DDR4-3200 1TB)\\\cline{1-1}
    \textbf{GPU} & 2$\times$ Nvidia A100 40GB PCIe with NVLink \\\cline{1-1}
    \textbf{SSD} & 7$\times$ Intel Optane P5800X 1.6TB. 2$\times$ RAID0 arrays. \\\cline{1-1}    \textbf{Software} &     \begin{tabular}[c]{@{}l@{}}Ubuntu 20.04.6 (5.15.0-113), CUDA 12.2 (driver\\535.183.01), PyTorch 2.2.2, DeepSpeed 0.14.2, \\Megatron-DeepSpeed~\cite{microsoftMicrosoftMegatronDeepSpeedOngoing2019} (latest), kvikio 24.08\end{tabular} \\
\bottomrule
\end{tabular}
}
\end{table}

\begin{figure}[!t]
\centering{\includegraphics[width=\linewidth]{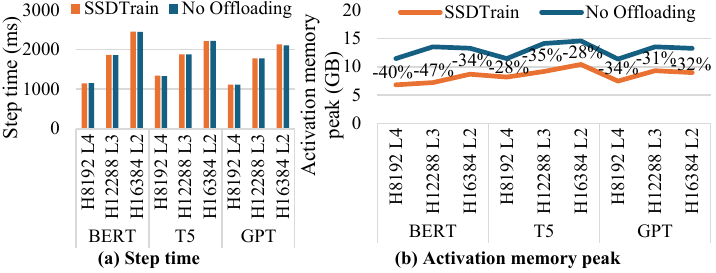}\vspace{-7pt}}
\caption{\label{fig:eval_perf} Comparing \oursystem{} to runs without tensor offloading. We test with different hidden dimensions (H) and number of layers (L). Batch size is 16.}
\end{figure}

We use the two GPUs for TP. The number of micro-batches per step is fixed at 1 because without PP, a new micro-batch will not start before both forward propagation and backward propagation of the previous micro-batch are done. \rv{More} micro-batch\rv{es} only brings in gradient accumulation and does not affect the activation offloading pattern. In other words, unless stated otherwise, the micro-batch size is equivalent to batch size throughout Sec.~\ref{sec:evaluation}. The hidden dimension is from 8192 to 16384, and we use typical hyperparameters~\cite{devlinBERTPretrainingDeep2019,touvronLlamaOpenFoundation2023,raffelExploringLimitsTransfer2023} for this range. The attention head dimension is 128. The text sequence length is 1024. For T5, the number of decoders is half of the total number of layers, rounded down. FlashAttention-2~\cite{daoFlashAttention2FasterAttention2023} is used with or without \oursystem{} for optimized attention computation.

As each A100 has only 40GB of device memory, to explore the design space closer to that in real-world training systems with A100 80GB and later GPUs~\cite{shoeybiMegatronLMTrainingMultiBillion2020a,liuWinnerTakeAllColumnRow2023}, we make several mitigations. First, we use FP16 instead of mixed precision, eliminating the FP32 weight copy. Second, we use SGD instead of Adam as the optimizer to reduce optimizer states. The two measures only affect accumulation operations and weight updates, thus imposing a constant bias in step time and memory use in execution with or without \oursystem{}.

\subsection{Performance and Peak Memory Usage}
\label{sec:exp_baseline}

To understand \oursystem{}'s impact on execution time and peak memory usage, we measure the step time of BERT, T5, and GPT and the memory peak during forward and backward propagation. The collected metrics of system with \oursystem{} and without are compared in Fig.~\ref{fig:eval_perf}. For each model, we collected three scenarios with different (hidden dimension, number of layers): (8192, 4), (12288, 3) and (16384, 2). As shown, \oursystem{} has almost no performance overhead in all cases. Although \oursystem{} and its optimizations introduce additional CPU logic, the performance comparison indicates that this logic is not on the critical path. Rather, GPU computation defines the critical path, and the CPU's role lies primarily in launching new GPU jobs before current GPU operations complete.  Thus, the CPU is underutilized, and \oursystem{}'s extra work does not lead to delay in new tasks reaching the GPUs. 
\oursystem{} effectively reduces the activations' memory use peak by 28\%--40\% in these cases.

\subsection{Comparing the Activations Placement Strategies}
\label{sec:exp_dse}

\oursystem{} opens up offloading activations to SSDs as an option besides keeping activations in the GPU memory, and activations checkpointing. We compare the three different strategies here on the recompute-offload-keep (ROK) curve\rv{:} 
Fig.~\ref{fig:eval_dse} shows the training of two 3-layer BERT models\rv{.} \rv{O}ne \rv{set the} hidden dimension \rv{as} 12K and the other \rv{set it} as 14K. In a ROK curve, each run is represented by a point. The x-axis is the activations memory peak. The y-axis is the model throughput~\cite{shoeybiMegatronLMTrainingMultiBillion2020a}, i.e., the number of algorithmic computations involved in the training step regardless of software and hardware implementation, e.g., whether the activations are recomputed, divided by the training step time.
In these two cases, \oursystem{} reduces the GPU activations memory peak, allowing for a larger batch size to attain higher throughput. Given the same batch size, \oursystem{} offloading attains the throughput the same as the throughput when the activations are kept in memory. Meanwhile, \oursystem{} gets a lower activations memory peak than the recomputation. Compared with keeping the activations in memory, \oursystem{} is able to double the batch size with the same activations memory budget.

Other than the three strategies, before FlashAttention~\cite{daoFlashAttentionFastMemoryEfficient2022}, Megatron~\cite{korthikantiReducingActivationRecomputation2022} proposed selective checkpointing that recomputes the core attention modules.
As we use FlashAttention, the core attention module is done in one kernel, eliminating these intermediate tensors. The effect of selective checkpointing with FlashAttention has negligible impact on performance and peak memory usage for activations.

\subsection{Discussion}
\label{sec:discussion}
\noindent
\textbf{Examining the modeling.} To understand the accuracy of the model in Sec.~\ref{sec:projected_life}, we compare \oursystem{}'s offloaded amount with the model estimate. As Table~\ref{tab:eval_activations} shows, the figures are close. We also compute the required PCIe write bandwidth, which is reduced when the hidden dimension gets larger. Typically, a model with more than 60B parameters has a hidden dimension of no less than 8K~\cite{touvronLlamaOpenFoundation2023,jordanhoffmannTrainingComputeOptimalLarge2022}. PCIe write bandwidth aligns with the estimate in Sec.~\ref{sec:projected_life}.

\begin{figure}[!t]
\centering{\includegraphics[width=\linewidth]{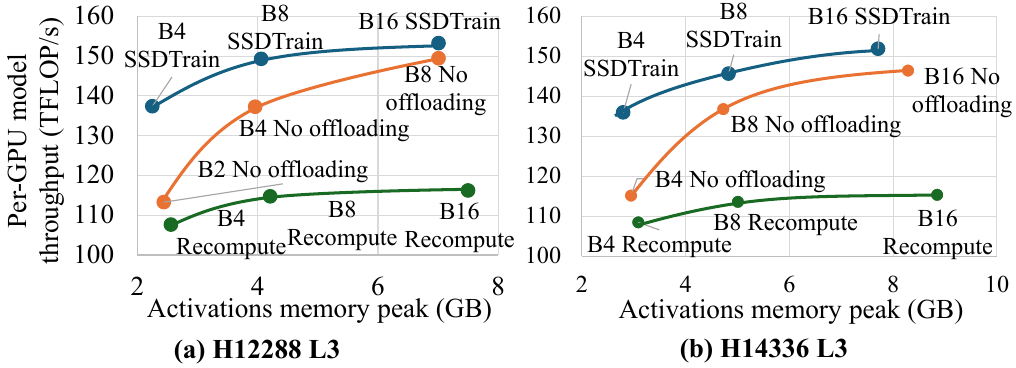}\vspace{-7pt}}
\caption{\label{fig:eval_dse}Recompute-offload-keep (ROK) curve of 3-layer (L) BERT with hidden dimension (H) as 12K or 14K. Different batch sizes (B) are tested.}
\end{figure}

\begin{figure}[!t]
\centering{\includegraphics[width=0.85\linewidth]{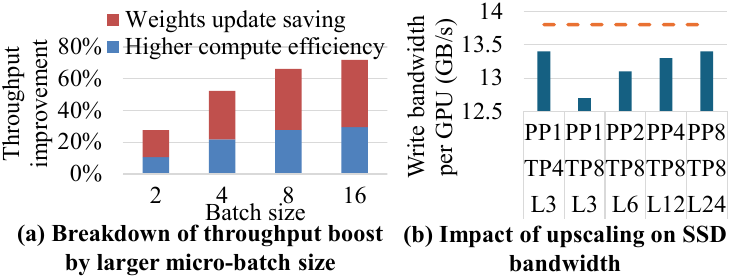}\vspace{-7pt}}
\caption{\label{fig:eval_case_study}Case study of 3-layer BERT with hidden dimension as 12K.}
\end{figure}

\begin{table}[!t]
{\footnotesize
\caption{The per-GPU offloaded tensor amount, estimate, and required PCIe write bandwidth of BERT with different hidden dimensions (H) and number of layers (L). Batch size is 16. \label{tab:eval_activations}\vspace{-5pt}}
\begin{tabular}{llll}
\toprule
                              & H8192 L4  & H12288 L3 & H16384 L2 \\\midrule
\textbf{Offloaded amount}     & 10.37 GB  & 12.85 GB  & 10.75 GB  \\
\textbf{Model estimate}       & 11.13 GB  & 12.60 GB   & 11.50 GB   \\
\textbf{PCIe write bandwidth} & 18.0 GB/s & 13.8 GB/s & 8.76 GB/s \\ \bottomrule
\end{tabular}
}
\end{table}

\noindent
\textbf{Impact of larger micro-batch size.} To further understand how larger micro-batch size improves the performance, we compare the no-offloading cases in Figure~\ref{fig:eval_dse}(a) to the same configurations with batch size as 1 and break down the throughput improvement in Fig.~\ref{fig:eval_case_study}(a). The improvement primarily comes from time saving by weights update, which is very relevant to large-scale LLM training systems. The micro-batch size is usually set small, e.g., 1 or 2 in Paxml~\cite{nitinScalingLargeLanguage2023} and BLOOM~\cite{workshopBLOOM176BParameterOpenAccess2023} pretraining, in exchange for smaller bubbles introduced by PP. In the BLOOM training system, each data parallel rank is assigned a mini-batch with 32 samples. When the micro-batch size is no less than 4, the ideal PP bubble time percentage is no less than 11.5\%. However, weight update and gradient accumulation cost is inversely proportional to the micro-batch size, which is huge when the micro-batch size is 1 or 2. SSDTrain allows larger micro-batch sizes given the same activation memory budget, thus beneficial to these PP-enabled training systems. 

\noindent
\textbf{Impact of upscaling.} 
Sec.~\ref{sec:llm_scaling} demonstrates that the whole-system activations size $S_{activations}$ grows slower than the whole-system GPU throughput $C$, i.e., $S_{activations} \propto C^{\frac{5}{6}}$. Therefore, the bandwidth required to fully overlap the computation with the SSD accesses is reduced.
In short, the scaling of LLM is essentially a weak scaling scenario, and the SSD IO latency is easier to hide when it is scaled up. 
In Fig.~\ref{fig:eval_case_study}(b), we further project the impact of upscaling on the write bandwidth per GPU using \texttt{llm-analysis}. We follow typical parallelism configurations~\cite{shoeybiMegatronLMTrainingMultiBillion2020a,nitinScalingLargeLanguage2023} when the number of GPUs is less than 100.  In all projected cases, the write bandwidth per GPU is smaller than the original 2-GPU case (orange dashed line). Vanilla DP only affects weights update and therefore has no effect on the write bandwidth. ZeRO may reduce the write bandwidth requirement due to the communication incurred in forward and backward propagation.

\noindent
\textbf{Cost analysis.} We study the SSD cost associated with adopting \oursystem{} offloading.
To get the endurance in Fig.~\ref{fig:projected_perf_model}, each A100 priced at US\$10K~\cite{dihuniNVIDIAA1009002100100000002021} is paired with in total US\$360 worth of SSDs.
The evaluation uses 7 Intel P5800X for the 2 A100s. Although P5800X is more expensive, the price per PBW is comparable~\cite{neweggIntelOptaneDC2021}.

\section{Related Work}

Many LLM systems with offloading abilities are inference-only~\cite{kwonEfficientMemoryManagement2023,shengFlexGenHighThroughputGenerative2023,alizadehLLMFlashEfficient2024}. In inference, weights and KV-cache never change and are reused across iterations; \rv{this is} leverage\rv{d} to enhance locality and memory efficiency. However, in training, all tensors, including the weights, change across the iterations. Some work avails offloading~\cite{rajbhandariZeROinfinityBreakingGPU2021} for training but is mostly designed to \rv{fit in} larger models at the cost of performance. \rv{A}sync transfer \rv{is missing} to \rv{keep} performance.
Another direction is to offload computation to the CPU~\cite{renZeROOffloadDemocratizingBillionScale2021,kamahoriFiddlerCPUGPUOrchestration2024,songPowerInferFastLarge2023}. The offloaded computation is light, and the offloaded data include gradients, sparse elements in the weights, etc. 
Our work is orthogonal: Interference with the CPU is minimized because we offload the activations to SSDs via GDS. Activations are for compute-intensive gradient computation, which is best done solely on GPUs. 

Before LLMs, there is work on offloading for deep learning~\cite{pengCapuchinTensorbasedGPU2020,wangSuperNeuronsDynamicGPU2018,baeFlashNeuronSSDEnabledLargeBatch2021,rhuVDNNVirtualizedDeep2016,huangSwapAdvisorPushingDeep2020}\rv{:} Most target main memory while some~\cite{baeFlashNeuronSSDEnabledLargeBatch2021} target SSDs. LLM is unique because massive parallelism and its \rv{memory} implications are fundamental to the design space. \oursystem{} naturally supports multiple GPUs. We showed its viability on clusters. Besides, LLM\rv{'s} demand for computing power \rv{is so high} that it stimulates rapid development in specialized hardware and frameworks. \oursystem{}{} ensures good interoperability, while most \rv{prior} work is bound to a specific PyTorch version or a custom runtime supporting select layers.

\section{Conclusion}

In LLM training, activations dominate the increasingly limited GPU memory. To address this, we propose \oursystem{} as an adaptive activation offloading framework to SSDs. We demonstrate its viability in large-scale systems by modeling. The evaluation shows \oursystem{} reduces the activations peak memory use by up to 47\% with negligible overhead. We analyze how this may lead to increased throughput by increasing micro-batch size and reducing pipeline bubbles.

\bibliographystyle{IEEEtran}
\bibliography{reference/ManuallyValidatedFlashtrainfromZotero,reference/manual}

\end{document}